\documentclass[prl,twocolumn,amsmath,floats,showpacs]{revtex4}

\usepackage{graphicx}
\usepackage{dcolumn}

\begin{document}
\title{Oxygen enhanced atomic chain formation}

\author{W.H.A. Thijssen}
\affiliation{Kamerlingh Onnes Laboratorium, Universiteit Leiden,
Niels Bohrweg 2, 2333 CA Leiden, Netherlands}

\author{D. Marjenburgh}
\affiliation{Kamerlingh Onnes Laboratorium, Universiteit Leiden,
Niels Bohrweg 2, 2333 CA Leiden, Netherlands}

\author{R.H. Bremmer}
\affiliation{Kamerlingh Onnes Laboratorium, Universiteit Leiden,
Niels Bohrweg 2, 2333 CA Leiden, Netherlands}

\author{J.M. van Ruitenbeek}
\affiliation{Kamerlingh Onnes Laboratorium, Universiteit Leiden,
Niels Bohrweg 2, 2333 CA Leiden, Netherlands}

\begin{abstract}
We report experimental evidence for atomic chain formation during
stretching of atomic-sized contacts for gold and silver, that is
strongly enhanced due to oxygen incorporation. While gold has been
known for its tendency to form atomic chains, for silver this is
only observed in the presence of oxygen. With oxygen the silver
chains are as long as those for gold, but the conductance drops
with chain length to about 0.1 conductance quantum. A relation is
suggested with previous work on surface reconstructions for silver
(110) surfaces after chemisorption of oxygen.
\end{abstract}

\date{\today}

 \maketitle

In recent years it has become possible to measure the electrical
and mechanical properties of the ultimate one dimensional metallic
conductor: a chain of single atoms. For gold, freely suspended
single-atom chains up to seven or eight atoms long have been
created and investigated \cite{yanson98,ohnishi98,smit03} and it
has been possible to identify its vibration modes by means of
point contact spectroscopy \cite{agrait02}. It has been argued
that the underlying physical mechanism for the formation of freely
suspended monatomic chains for Au, Pt, and Ir is the same as for
the surface reconstructions on clean [110] surfaces of these 5d
metals \cite{smit01}. The 4d metals that are positioned above
these 5d metals in the periodic system, such as Ag in the case of
Au, were found in low temperature break junction experiments to
form only short chains, just two, or rarely three atoms long
\cite{smit01,Roelthesis}. Indeed Ag does not show the same
tendency to surface reconstructions, but when the [110] surface
reacts with oxygen it does reconstruct into an added row
reconstruction with -Ag-O-Ag- rows lying in the [001] direction on
the surface. \cite{zanazzi,canepa}

Previous transmission electron microscopy (TEM) measurements on
suspended Au chains, created at room temperature and under
ultra-high vacuum (UHV) conditions have shown anomalously large
interatomic distances of up to 4 \AA\  \cite{ohnishi98}, compared
to 2.9 \AA\ in bulk Au. On the other hand, Au atomic wires created
under cryogenic vacuum at 4.2 K reveal an interatomic distance of
$2.5\pm 0.2$ \AA\  \cite{untiedt02}, which is consistent with
calculations on the formation of Au wires
\cite{hakkinen00,bahn01,dasilva01}. By a similar room temperature
TEM method atomic Ag chains with large interatomic distances have
been observed as well\cite{rodrigues02} which is at variance with
the low temperature break junction results referred to above
\cite{Roelthesis}. It has therefore been suggested that the
observation of these large interatomic distances can be attributed
to the incorporation of small foreign atoms into a Au or Ag chain,
which cannot be imaged by TEM
\cite{legoas02,novaes03,skorodumova03}. Recent density functional
theory calculations have predicted that oxygen can be stably
incorporated into a Au chain \cite{bahn02}. The Au-O bond was
calculated to be stronger than a Au-Au bond, as seen from a
lowering of the total energy of a Au-O chain compared to a Au
chain, which would imply that also longer chains could be formed.
The conductance of these Au-O chains should stay close to one
conductance quantum G$_{0}$ = 2e$^{2}$/h, as is the case for a
pure Au chain, which means that the chain will have an almost
perfect transmission for electrons. In this letter we discuss
experiments on admitting oxygen to atomic chains of Au and Ag
atoms. We present evidence that oxygen is chemically reacting with
the atomic chains, is being incorporated thus stabilizing the
atomic chains. For Ag the presence of oxygen allows the formation
of chains as long as those for Au. Our observations open the way
to investigate mixed atomic wires with inclusions of foreign atoms
or molecules.

The experiments were done using the mechanically controllable
break junction technique (see Ref.~\cite{agrait03} for a detailed
description). The sample wire is first broken under cryogenic
vacuum at 4.2 K, in order to create clean fracture surfaces. The
temperature of the sample can be varied using a local heater and
thermometer, while the vacuum can remains at 4.2 K. O$_2$ can be
admitted via a capillary, equipped with a heater wire running all
along its interior that prevents premature condensation of the
gas.

\begin{figure} [t!]
\includegraphics[width=8cm,angle=0] {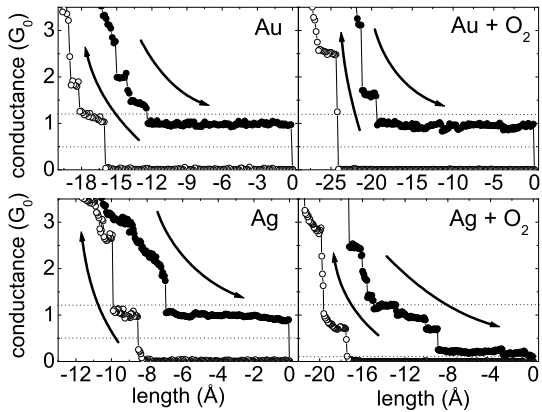}
\caption{\label{traces} Individual breaking and return traces for
clean Au, for Au after admitting O$_2$, for Ag, and for Ag after
admitting O$_2$. A bias voltage of 50 mV was used in all cases.}
\end{figure}

\begin{figure} [b!]
\includegraphics[width=8cm,angle=0]{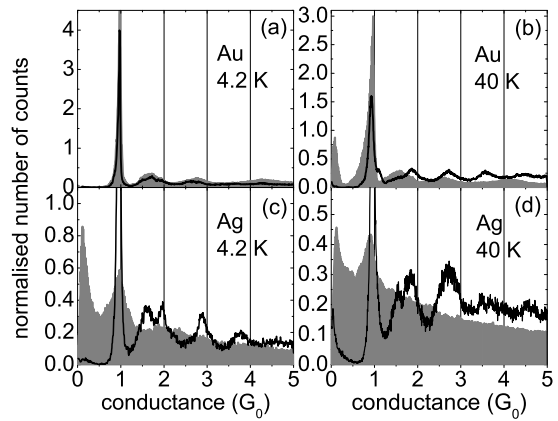}
\caption{\label{conductance} Conductance histograms for (a) Au
(black curve) and Au after admitting O$_2$ (filled graph) at 4.2
K; (b) Au (black curve) and Au after admitting O$_2$ (filled
graph) at 40 K; (c) Ag (black curve) and Ag after admitting O$_2$
(filled graph); (d) Ag (black curve) and Ag after admitting O$_2$
at 40 K. Due to scaling the 1 G$_{0}$ peak maxima for clean Ag in
(c) and (d) are not seen; maximum values are 2.2 and 1.8
respectively. A bias voltage of 50 mV was used and each graph is
build from approx. 2000 breaking traces.}
\end{figure}

When a Au or Ag wire is broken gently, a thin neck is being formed
between the two contact leads. By further stretching of the
contact the neck is thinned and finally a contact of only a single
atom is formed. During the final stages of breaking one observes a
step-like decrease of the conductance and a final last plateau,
that has a conductance near 1 G$_{0}$. Figure~\ref{traces} shows
several typical conductance traces for clean Au, for Au after
admitting O$_2$, for clean Ag, and for Ag after admitting O$_2$.
The last plateau of a breaking trace for Au can often be stretched
to quite long lengths, which signals the formation of monatomic
chains \cite{yanson98}.
By collecting a large set of these traces and monitoring the
conductances during the breaking process a conductance histogram
can be constructed. Both for Au and Ag at 4.2 K the conductance
histogram is dominated by a single sharp peak at 1 G$_0$,
indicating that a contact of a single atom in cross section is
occurring frequently, Fig.~\ref{conductance}. Admitting O$_2$ to
Au and Ag atomic contacts at 4.2 K results in clearly different
conductance histograms for the two metals: The conductance in the
case of O$_2$ admission to Au does not significantly change. But
when O$_2$ is admitted to a Ag atomic contact a dominant peak
around 0.1 G$_{0}$ appears together with a large background,
indicating stable configurations with conductances below 1 G$_0$
are regularly formed, Fig.~\ref{conductance}(c). As the histograms
have been normalized to the area under the curve one can see that
the weight of the peak at 1 G$_{0}$ has been transferred to the
structure at lower conductances. When clean Au is heated to 40 K
the average chain length decreases, which can be seen from the
lower relative weight of the peak at 1 G$_{0}$ in
Fig.~\ref{conductance}(b). However, when O$_2$ is admitted the
peak around 1 G$_{0}$ becomes more prominent and in addition a
peak at 0.1 G$_{0}$ appears. On the other hand, Ag with O$_2$ at
40 K shows a structure similar to the histogram at 4.2 K but with
further increased background.

\begin{figure} [t!]
\includegraphics[width=8cm]{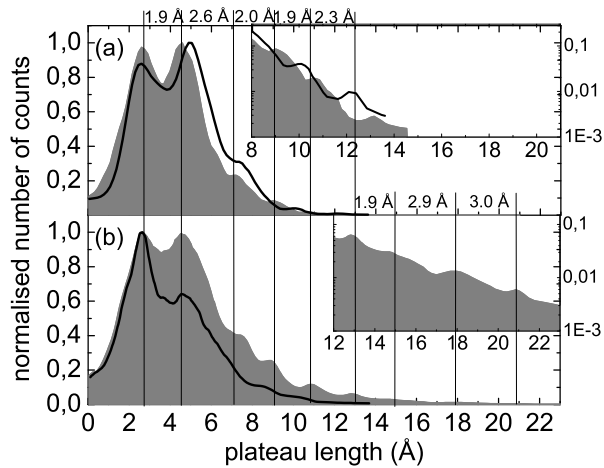}
\caption{\label{lhisto} Figure (a) shows a length histogram for Au
(black curve) and for Au after admitting O$_2$ at 4.2 K (filled
graph); figure (b) shows a length histogram for Au (black curve)
and for Au after admitting O$_2$ (filled graph) at 40 K. The
insets of both figures show semi-log plots of the tails of the
distributions in order to bring out the small peaks. Vertical
lines are shown at the peak positions in the length histograms for
Au with oxygen, together with their distances. In all cases a bias
voltage of 50 mV was used and the start and stop conductance
values between which the lengths of the plateaus were measured,
were 1.1 G$_{0}$ and 0.5 G$_{0}$ respectively}
\end{figure}

It is instructive to analyze the lengths of the chains that are
being formed for the case of Au and Ag, with and without oxygen
admission. By repeating the process of breaking and making a
contact several thousands of times and automatically recording the
lengths of the last plateaus near 1 G$_{0}$ one can construct a
length histogram. The length starts to be recorded when the
conductance drops below 1.2 G$_{0}$ and stops for G $<$ 0.5 G$
_{0}$. A typical length histogram for Au recorded at 4.2 K,
reveals several equidistant peaks that decrease in height
exponentially after the first two peaks, as can be seen from the
black curve in Fig.~\ref{lhisto}(a). The distance between the
peaks gives the interatomic distance of the atoms in the chain
\cite{yanson98,untiedt02}.  Numerical simulations show that the
conductance of a Au atomic contact drops just below 1 G$_{0}$ when
both contact apexes end in a single atom
\cite{dreher,palacios02,jelinek04}. Starting from that
configuration the length of the chain is being recorded in our
experiments. The position of the first peak in the histogram can
thus be interpreted as being the difference in length between a
one atom contact and a chain of two atoms, both at the point of
breaking. The position of the second peak is the difference in
length between a one atom contact and a chain consisting of three
Au atoms etc. The histogram shows that chains of two and three
atoms long have a high probability to be formed and that the
probability for the formation of longer chains decreases very
rapidly. Chains more than seven atoms long are very exceptional.
After O$_2$ is admitted to the contact region and the contact
itself is kept at 4.2 K, a small but clearly observable change in
the distance between some peaks in the length histogram is
observed, Fig.~\ref{lhisto}(a). Some of the peaks of the filled
graph (Au with O$_2$) are shifted somewhat to the left compared to
the peaks of the black curve (clean Au), notably the second peak
and the peaks that are made visible the inset of
Fig.~\ref{lhisto}(a). A closer look at the distances in
figure~\ref{lhisto}(a) observed for Au after admitting O$_2$,
suggests two sets of distances, namely 2.5 $\pm$ 0.2 \AA, seen
e.g. between the second and third peak, and 1.9 $\pm$ 0.2 \AA,
seen e.g. between the first and second peak.
When the temperature of a clean Au contact is raised to 40 K a
faster decrease of the intensity of the peaks is seen, as the
result of thermal decay of the metastable chain configuration.
Figure~\ref{lhisto}(b) shows a length histogram for clean Au at 40
K (black curve). When at 40 K O$_2$ gas is admitted a remarkable
observation is made. Again the distances between some peaks change
but the average length of the chains is significantly longer, even
longer than at 4.2 K. Chains up to ten atoms long are observed, as
seen in Fig.~\ref{lhisto}(b).
\begin{figure} [b!]
\includegraphics[width=8cm,angle=0]{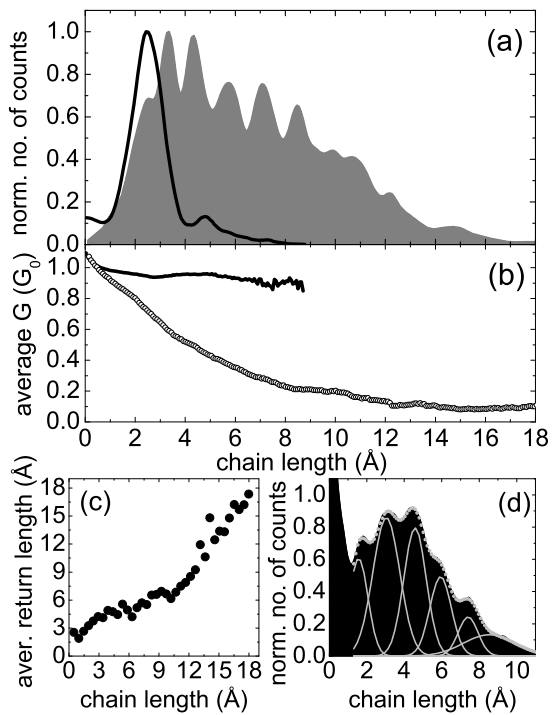}
\caption{\label{lhistoag}(a) Length histogram of clean Ag (gray
curve) and a length histogram of Ag after admitting O$_2$ (filled
graph) at 4.2 K; (b) Average conductance as a function of chain
length for clean Ag (black curve) and Ag with admitted O$_2$
(dotted curve); (c) Average return lengths as a function of Ag-O
chain lengths; (d) Length histogram of Ag after admitting O$_2$
with start and stop values of 0.5 G$_{0}$ and 0.05 G$_{0}$
respectively.}
\end{figure}

Even more spectacular results were obtained when we performed
these same experiment on Ag. A length histogram for Ag recorded at
4.2 K is shown as the black curve in Fig.~\ref{lhistoag}(a). O$_2$
was admitted when the Ag atomic contact was at 4.2 K. Since a peak
around 0.1 G$_{0}$ appears (see Fig.~\ref{conductance}) the start
and stop conductance values for measuring the chain length were
put at 1.2 G$_{0}$ and 0.05 G$_{0}$ respectively. It is very
striking that the average chain length is dramatically increased
compared to a clean Ag contact. A sequence of peaks is observed
indicating the repeated occurrence of certain stable chain
configurations. When a chain of single atoms breaks its
conductance usually drops to a very low conductance, deep into the
tunnelling regime. This is expected because after a chain breaks,
the atoms collapse back to the leads resulting in a large tunnel
distance between the electrodes. In the lower right panel of
Fig.~\ref{traces} a typical breaking trace and its return trace
for Ag with oxygen are shown. The return trace shows the
conductance of the contact when it is being made again after it
has been broken. The observation of additional plateaus with
conductance values as low as 0.1 G$_{0}$ indicates that the chain
is still mechanically intact. We have verified that the
conductance plateaus are indeed due to chain formation by
measuring the average return length as a function of the length of
the chains. We confirm that these two lengths are, on average,
nearly the same, see Fig.~\ref{lhistoag}(c).
Figure~\ref{lhistoag}(b) shows the average conductance of the
clean Ag chains and for Ag-O chains as function of chain length.
The curves are obtained by adding all measured conductance traces
from the start value (1.2 G$_0$) onward, and dividing at each
length by the number of traces included at that point. For clean
Ag the conductance remains close to 1 G$_0$, with small parity
oscillations with the number of atoms in the chain, as was
previously observed for Au \cite{smit03}. In contrast, the average
conductance for Ag-O chains decreases rapidly as the chains become
longer, stabilizing at approximately 0.1 G$_{0}$. Further evidence
for chains incrementing by discrete atomic units at lower
conductances can be found from Fig.~\ref{lhistoag}(d) which shows
a length histogram for Ag with O$_2$ for which the start and stop
values were set to 0.5 G$_{0}$ and 0.05 G$_{0}$, respectively. At
least five peaks can clearly been seen and the histogram is
perfectly fitted with a multiple gaussian that gives an inter peak
distance of again 1.5 $\pm$ 0.2~\AA. It is worth mentioning that
similar results were obtained when these experiments on Ag-O chain
were done at 40 K.

In order to interpret these unexpected results we consider the
possibility of atomic oxygen being incorporated into Au and Ag
chains. It is known that low-coordinated Au atoms are chemically
very reactive \cite{valden98,Landman99}. As Au atoms have a
coordination number of two in a chain an atomic chain is expected
to be even more reactive then a nanoparticle, opening the
possibility for molecular oxygen to dissociate even at low
temperatures. The oxygen atoms are then expected to be
incorporated in the Au chain, as predicted by Bahn {\it et al.}
\cite{bahn02}. We have seen that the length-histogram peak
distances change upon admitting O$_2$ gas. The distance of 1.9
\AA\ agrees with the Au-O bond in the calculations \cite{bahn02}.
At 40 K considerably longer chains are formed after admitting
O$_2$, which agrees with the Au-O bond being stronger than the
Au-Au bond \cite{bahn02}. We expect that at 4.2 K an O$_2$
molecule can only be dissociated when it lands directly on the
atomic chain. O$_2$ molecules that land on the leads will remain
frozen at the landing spot and will not dissociate because higher
coordinated Au atoms are less reactive. The fact that peaks are
observed in the length histogram suggests that specific chain
compositions are preferentially formed. The observed chain
compositions can be explained by assuming only one O$_2$ molecule
that has dissociated on the Au chain, suggesting that at 4.2 K
only two oxygen atoms are typically incorporated in the chain.
When the temperature is increased to 40 K the vapor pressure of
O$_2$ increases significantly and the mobility on the surface is
enhanced. This may create a sufficient amount of oxygen atoms for
building a long chain, which is reinforced relative to a clean Au
chain.

In the case of Ag with oxygen we observe a distance between peaks
of 1.5 $\pm$ 0.2~\AA\ and long chains even at 4.2 K, suggesting
that low-coordinated Ag atoms dissociate O$_2$ molecules more
readily. The first dominant peak of the clean Ag length histogram
coincides with the shoulder at small lengths of the histogram for
Ag with oxygen. This suggests that the first true peak for the
latter is a Ag-O-Ag chain and that the shoulder is a Ag-Ag chain,
which is less frequently formed.

We conclude that oxygen atoms are incorporated in chains of Au or
Ag atoms, thus reenforcing it, which is most apparent in the case
of Ag. We suspect a fundamental connection between the mechanism
for oxygen induced surface reconstructions on Ag(110)
\cite{zanazzi,canepa} and the strongly enhanced chain formation.
The fact that oxygen chemisorption leads to similar metal-oxygen
row reconstructed surfaces for other metals opens possibilities
for further investigation of enhanced chain formation due to
oxygen incorporation. Preliminary results for Cu show indeed that
oxygen also induces long chain formation. The mere observation
that foreign atoms and molecules can chemically react locally at
an atomic contact and even be incorporated into an atomic chain
opens new possibilities for modifying and investigating different
combinations of metal contacts and chains with foreign molecules
and atoms. It may even be possible in the future to exploit
atomically thin wires to contact single atoms or molecules which
would provide ideal quantum leads to the nano-object of study.

We thank P. Jel{\'\i}nek and F. Flores for valuable discussions
and for communicating unpublished work. This work is part of the
research program of the ``Stichting FOM,'', partially financed
through the SONS programme of the European Science Foundation,
which is also funded by the European Commission, Sixth Framework
Programme, and was also supported by the European Commission TMR
Network program DIENOW.

\end{document}